\documentclass{appolb}
\usepackage{graphicx} 
\usepackage{amsmath}

\title{Entropy and mean multiplicity from dipole models in the high energy limit}
\author{Krzysztof Kutak$^{1,2}$, S\'andor L\"ok\"os$^{1,3}$
\address{$^1$Institute of Nuclear Physics Polish Academy of Sciences \\
$^2$CPHT, CNRS, Ecole Polytechnique, Institut Polytechnique de Paris \\
$^3$Hungarian University of Agriculture and Life Sciences, Institute of Technology}}
\date{March 2026}

\begin{document}

\maketitle
\begin{abstract}
The 1D Mueller dipole model, its high energy limit, and its generalization were investigated. To address the ambiguity stemming from different definitions of the pseudorapidity ranges in experimental measurements, we propose the entropy as the function of the logarithm of the average multiplicity, $S(\ln\langle n\rangle$, as a universal observable. From the solutions of the models, we calculate both the entropy and the average charged particle multiplicity and compare to data measured in proton-proton collisions. We obtained these quantities directly from the measured charged particle multiplicity distributions and determine the model parameters via fits. We find that the generalized dipole model provides a significantly better description of the data than the 1D Mueller model.
\end{abstract}

\section{Introduction}

Charged particle multiplicity distributions have been measured since the dawn of experimental particle physics. An increasing interest is observed nowadays in such data due to recent theoretical proposal that link the Shannon's entropy that is associated to the multiplicities, to the entanglement entropy of the initial partonic system. It is conjectured that the initial state in the high energy limit is maximally entangled, therefore the associated entanglement entropy is maximal and can be expressed as $S_\textmd{init}=\ln\langle n\rangle$, where $\langle n\rangle$ is the average number of partonic microstates. This conjecture was investigated in deep inelastic scattering experiments with the H1 detector \cite{H1:2020zpd} and in recent theory calculations based on parton distribution functions \cite{Hentschinski:2021aux,Hentschinski:2022rsa,Hentschinski:2023izh}. Good agreement was found between the calculations and the data.

Describing the multiplicities in the final state, maximum entropy principle \cite{PhysRev.106.620,PhysRev.108.171,Lokos:2025cbu} also can be utilized. It was shown, however, that the shape of the distributions is different for the $\eta$ pseudorapidity variable definitions adopted by HERA \cite{H1:2020zpd} and LHCb \cite{LHCb:2014wmv}, where a narrow $\Delta\eta$ window is shifted across the acceptance, and that is adopted by ALICE \cite{ALICE:2010cin,ALICE:2015olq,ALICE:2017pcy}, ATLAS \cite{ATLAS:2010jvh,ATLAS:2016zkp} CMS \cite{CMS:2010qvf} and UA5 \cite{UA5:1988gup}, where the rapidity window with increasing width is centered to $\eta=0$. Such difference requires extra attention when one compares model calculations or measurement results.

Our proposition \cite{Kutak:2025tsx} is to use the universal entropy function $S(\ln\langle n \rangle)$ to compare measured entropies from different experiments and model calculations. This observable is straightforward to determine from the experimental data from their definition
\begin{align}
    \langle n \rangle {=} \sum_n n P(n) \hspace{0.3cm}{\rm and }\hspace{0.3cm} S_h {=}{-}\sum_n P(n) \ln(P(n)).
\label{eq:basic_defs}
\end{align}
We obtain the universal $S(\ln\langle n \rangle)$ from charged particle multiplicity distribution measured in $p+p$ collisions \cite{LHCb:2014wmv, UA5:1988gup, ALICE:2015olq, ALICE:2017pcy, ATLAS:2010jvh,ATLAS:2016qux,ATLAS:2016zkp, CMS:2010qvf} and compare them to model calculations such as the 1D Mueller dipole model \cite{Mueller:1994gb,Levin:2003nc,Kharzeev:2017qzs} and one of its extension \cite{Caputa:2024xkp}.

\section{The dipole cascade models}

The utilized dipole models are, the 1D Mueller model \cite{Mueller:1994gb,Levin:2003nc,Kharzeev:2017qzs} and its generalization \cite{Caputa:2024xkp,Kutak:2025tsx} that describe parton cascades that develops, in the 1D case, in rapidity. The generalized model has an additional parameter that allows for more accurate description of multiplicity.

\subsection{1D Mueller model}

The 1D Mueller model that depends only on rapidity is given with the following cascade equation
\begin{align}
     \partial_y P_n(y) = -\alpha n P_n(y) + (n-1)\alpha P_{n-1} (y) ,
\end{align}
where the $P_n(y)$ is the initial state parton multiplicity distribution and $\alpha$ represents the gluon emission kernel and characterizes the speed of the Balitsky--Fadin--Kuraev--Lipatov (BFKL) cascade evolution \cite{Balitsky:1978ic,Kuraev:1977fs}. The solution of the equation can be written in the form of
\begin{align}
    P_n(y) = \frac{1}{C}e^{-\alpha y} \left( 1 -\frac{1}{C}e^{-\alpha y} \right)^{n-1},
\label{eq:geom_solution}
\end{align}
with the recently introduced $C$ parameter \cite{Hentschinski:2022rsa}. One can observe that, by introducing the probability $p=\frac{1}{C}e^{-\alpha y}$, the solution has the form of the geometric distribution, therefore the mean multiplicity can be expressed as
\begin{align}
    \langle n \rangle = \frac{1}{p} = Ce^{\alpha y} \hspace{0.5cm} \Rightarrow \hspace{0.5cm} y=\frac{1}{\alpha}\ln\left( \frac{\langle n \rangle}{C} \right).
\end{align}
and, by applying Eq. (\ref{eq:basic_defs}), a relation between the mean multiplicity and the entropy \cite{Kharzeev:2017qzs} can be obtained as:
\begin{align}
    S(y) = \langle n \rangle \ln \langle n \rangle - (\langle n \rangle - 1)\ln(\langle n \rangle - 1) \approx \ln(\langle n \rangle)+1.
\label{eq:Sgeom}
\end{align}
where in the last step, we took the $C=1$ in the sub-asymptotic limit \cite{Hentschinski:2023izh}.

\subsection{Generalized Mueller model}

A new cascade equation that was introduced in Ref.~\cite{Caputa:2024xkp} has a new parameter as
\begin{align}
    \partial_y P_n(y)=-\alpha (n+2h) P_n(y) + \alpha (n-1+2h)P_{n-1} (y)
\label{eq:CK}
\end{align}
and its solution has the form of the negative binomial distribution with the above defined $p=\frac{1}{C}e^{-\alpha y}$ probability:
\begin{align}
    P_n(y)=\frac{\Gamma(2h+n)}{n!\Gamma(2h)} p^{2h} \left( 1-p\right)^{n}.
\label{eq:NBD_solution}
\end{align}
The equation has been obtained in the context of study of Krylov complexity \cite{Nandy:2024evd,Rabinovici:2025otw,Baiguera:2025dkc} for the coherent states of the conformal SL(2,R) group and mapped to QCD in Ref.~\cite{Caputa:2024xkp}. Here $h$, is a conformal weight of this group which tells how the operators of the theory scale. As an example, in the previous section, we described the Mueller model that has the conformal weight $h=1/2$.

In our studies it also measures the deviation from the Mueller case and the dispersion of the NBD as $k_{NBD}=2h$. It is also worth to emphasis that we utilize a version of the model that allows for $p_0$, which effectively corresponds to the vacuum contribution. Such contributions were accounted for in recent studies \cite{Hentschinski:2024gaa,Kutak:2025syp} and successful description of hadronic entropy data were obtained.

An explicit form of the entropy of this distribution is not known (an implicit, closed expression is given in Ref.~\cite{DBLP:journals/corr/abs-1708-06394}). The mean multiplicity of dipoles or, in the context of Ref.~\cite{Caputa:2024xkp}, Krylov complexity (that is the measure of the spread of the underlying quantum state in Hilbert space), as the function of rapidity in the following form
\begin{align}
    \langle n\rangle= 2h \left(C e^{\alpha y}-1 \right).
\label{eq:navg_y_NBD}
\end{align}

\subsection{Comparison of the models}

\begin{figure}
    \centering
    \includegraphics[width=0.75\textwidth]{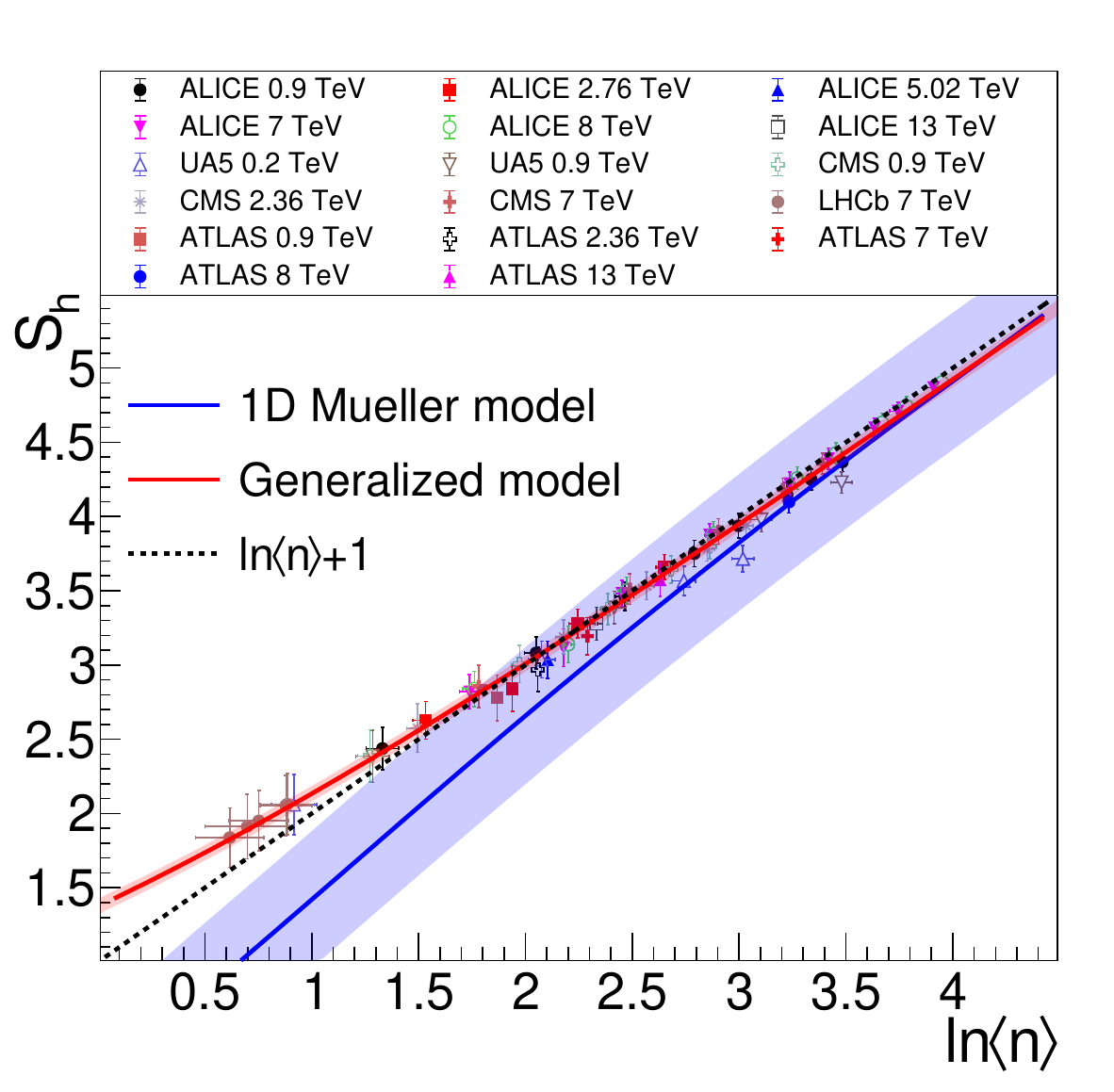}
    \caption{The new observable compared to data}
    \label{fig:S_lnnavg}
\end{figure}

We show the comparison of the two studied models in terms of logarithm of average multiplicity and entropy. The models are evaluated with parameters based on fits that will be discussed in Sec.~\ref{sec:datacomparison}. From the solutions given in Eqs.~\eqref{eq:geom_solution} and \eqref{eq:NBD_solution}, the multiplicity distributions for the original dipole model features exponential decay, the distributions from the generalized model are more disperse that closely resemble the measured multiplicity data in $pp$ \cite{LHCb:2014wmv, UA5:1988gup, ALICE:2015olq, ALICE:2017pcy, ATLAS:2010jvh,ATLAS:2016qux,ATLAS:2016zkp, CMS:2010qvf} and $ep$ \cite{H1:2020zpd}. It can be seen in Fig.~\ref{fig:S_lnnavg} that the generalized model predicts higher average number of partons and higher entropy in the investigated rapidity range.

\begin{figure}
    \centering
    \includegraphics[width=0.49\textwidth]{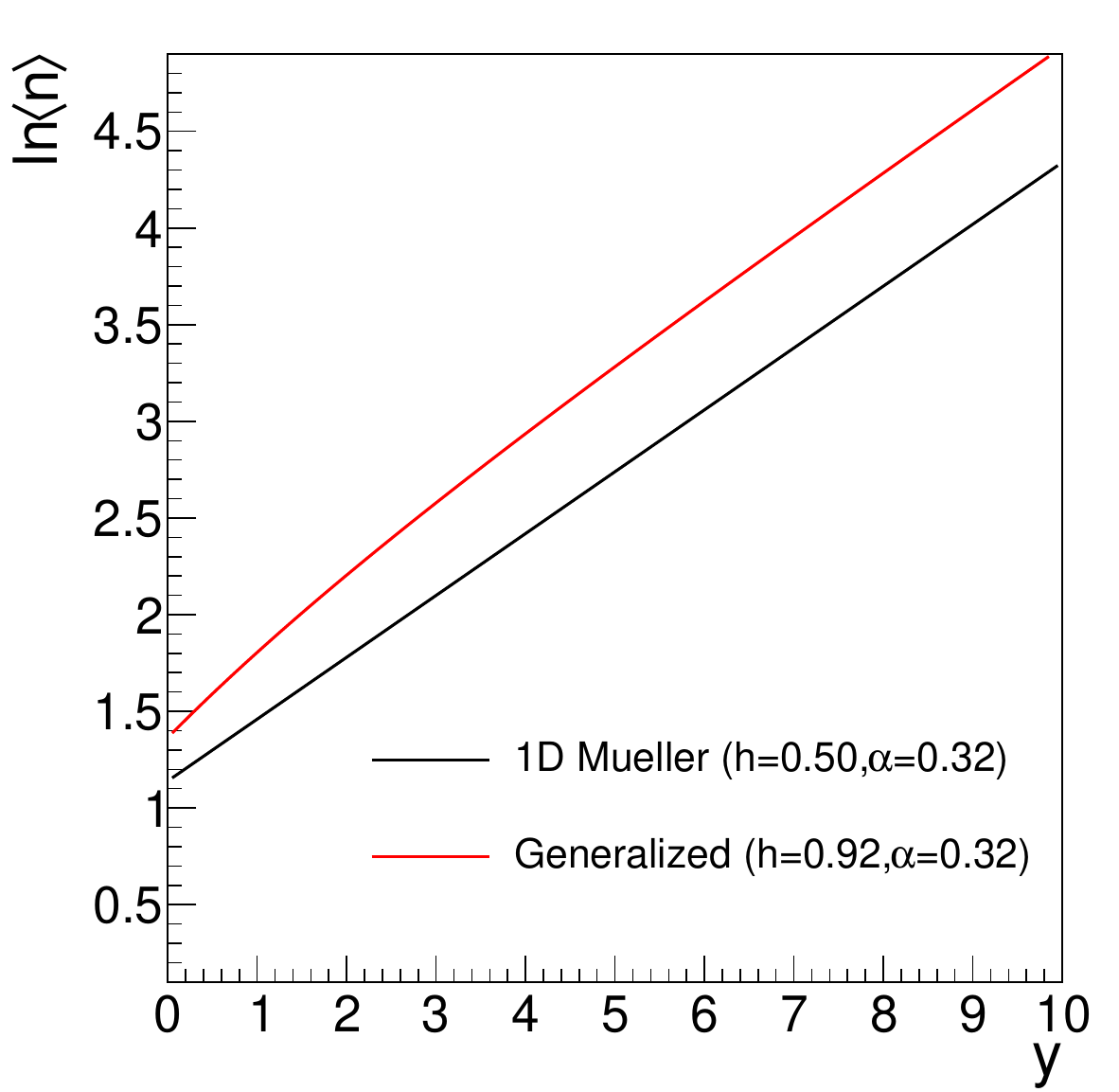}
    \includegraphics[width=0.49\textwidth]{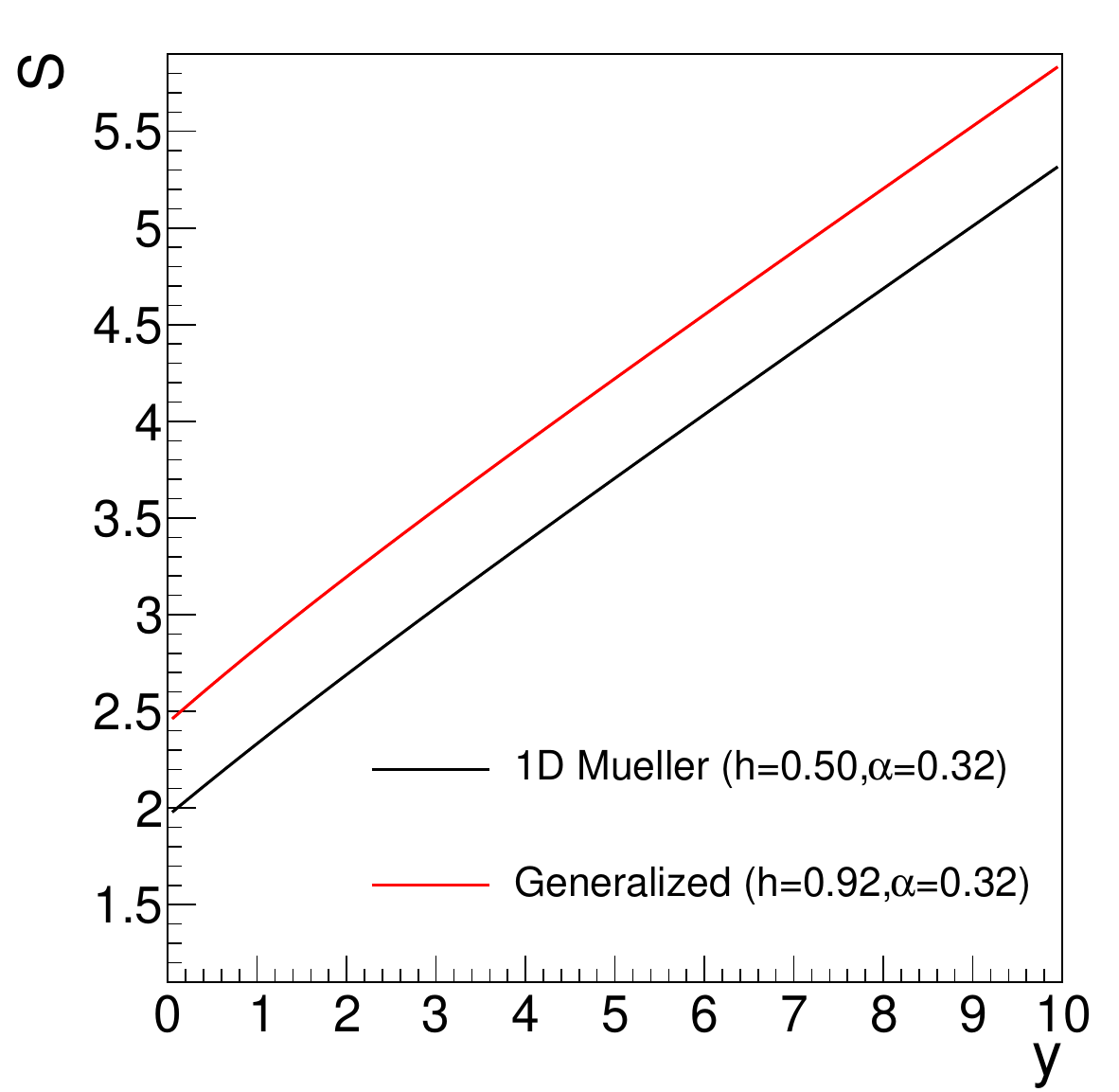}
    \caption{The log average multiplicity (left) and the entropy (right) as the function of the rapidity calculated from the two models.}
    \label{fig:model_comparison}
\end{figure}

\section{Comparison to data}
\label{sec:datacomparison}

We calculated the $S(\langle n \rangle)$ from both models and compared the results to data from $pp$ collision ranging from $\sqrt{s_{\textmd{NN}}}=$200 GeV to 13 TeV, from $|\eta|<0.5$ to $4.5<\eta<5$ in pseudorapidity \cite{LHCb:2014wmv, UA5:1988gup, ALICE:2015olq, ALICE:2017pcy, ATLAS:2010jvh,ATLAS:2016qux,ATLAS:2016zkp, CMS:2010qvf}. We found that the generalized model can well describe the data, however, the Mueller model exhibits deviation from the data at lower multiplicities. The obtained model parameters and the goodness of the fits can be seen in Tab.~\ref{tab:CK_data_comparison}.

\begin{table}[]
    \centering
    \begin{tabular}{|c|c|c|}
    \hline
       Parameter  & 1D dipole & Generalized model \\
    \hline
       $h$      & 0.50 (fixed)  & $0.92\pm0.05$  \\
       $\alpha$ & 0.32 (fixed)  & 0.32 (fixed)  \\
       $C$      & $3.13\pm0.48$ & 3.13 (fixed) \\
       \hline
       $\chi^2/$NDF & 309/63   & 24/63 \\
    \hline
    \end{tabular}
    \caption{The parameters and fit qualities of the dipole models \cite{Mueller:1994gb,Levin:2003nc,Kharzeev:2017qzs,Caputa:2024xkp}. The fixed values are based on earlier results \cite{Hentschinski:2024gaa}.}
    \label{tab:CK_data_comparison}
\end{table}

\section{Summary and outlook}

We described the 1D Mueller dipole model and its generalized version. We introduced a new observable, $S(\ln \langle n \rangle)$, that has been compared to $pp$ data spanning 2 orders of magnitude in energy and covers a large pseudorapidity interval. We found that the generalized model can describe the data while the Mueller model deviates from the data points at lower multiplicities.

We would like to emphasis that such studies could be valuable in the context of saturation if the model would be extended to take into account for recombination effects. Also, theoretical calculation could improve the description of $pp$ data as more complex colliding environment than DIS processes. Experimentally, the addressed methodological issue could be resolved and further, more detailed analyses could refine the present and future results.

\bibliographystyle{prlsty}
\bibliography{main.bib}

\end{document}